\documentclass[twocolumn,twoside,slac_two]{revtex4}
\usepackage{graphicx}
\usepackage{fancyhdr}
\begin{document}

\title{Looking for Signs of Anisotropic Cosmological Expansion in the 
\\ High-z Supernova Data}

\author{Brett Bochner}
\affiliation{Hofstra University, Hempstead, NY 11549, USA}

\begin{abstract}
Several problematical epochs in cosmology, including the recent period 
of structure formation (and acceleration), require us to understand 
cosmic evolution during times when the basis of FRW expansion, the 
cosmological principle, does not completely hold true. We consider that 
the breakdown of isotropy and homogeneity at such times may be an 
important key towards understanding cosmic evolution. To study this, 
we examine fluctuations in the high-\textit{z} supernova data to search 
for signs of large-scale anisotropy in the Hubble expansion. Using a 
cosmological-model-independent statistical analysis, we find mild 
evidence of real anisotropy in various circumstances. We consider 
the significance of these results, and the importance of further 
searches for violations of the cosmological principle.
\end{abstract}

\maketitle

\thispagestyle{fancy}

\section{Introduction and Research Rationale}
After a wave of successes in recent years, cosmology today is facing 
a series of potentially major turning points. The accepted standard 
model of the field, the concordance model~\cite{refConcord}, is a 
patchwork of detailed and observationally proven theories, interlaced 
with paradigms that are equally well accepted, though less well proven 
and far less detailed in terms of precise physical models. In the near 
future, new data will either bring these paradigms dramatically into 
focus, or will force great changes into the tightly interwoven 
concordance model.

Besides the unsolved problem of baryogenesis, the two most problematical 
epochs in cosmology both involve similar circumstances: they seek to 
understand the Friedmann-Robertson-Walker (FRW) expansion of the 
universe during phases in which the fundamental conceptual basis of 
FRW expansion, the Cosmological Principle (i.e., isotropy and 
homogeneity)~\cite{refKT}, does not completely hold true. The early 
such epoch (the pre-homogenization period) is usually handled with 
the Inflation paradigm~\cite{refKT}, and the late such epoch (post-CMB 
formation of large scale structure) is currently addressed with 
paradigms such as Vacuum Energy~\cite{refKT}, 
Quintessence~\cite{refQuint} (e.g., Tracker 
Quintessence~\cite{refTrkQnt}), etc. Correct or not, these paradigms 
are not yet well constrained or proven in detail; and we take the 
position that it is no coincidence that the two least well understood 
epochs in cosmology happen to occur at times in which the Cosmological 
Principle loses its validity.

Demonstrating any direct theoretical link between the violation of 
isotropy and homogeneity, and solutions to specific cosmological 
problems, is beyond the scope of this current analysis; nor do we 
include here any investigation of the pre-homogenization period in 
the very early universe. What we focus upon here is the late epoch 
during which the CMB-era smoothness finally breaks down into clumpy 
structure. We seek to determine how seriously the Cosmological 
Principle is broken in the recent universe, and the tool we use for 
this purpose is the collection of high-redshift Type Ia supernova 
data used for measuring the cosmic acceleration 
(e.g., \cite{refRiess1}), representing perhaps the best mapping of 
the expansion flow on very large scales. Measuring the extent of any 
irregularity in the smooth Hubble expansion would be a key step 
towards determining its significance (if any) in our understanding 
of cosmological concordance.

Theoretically, there remains room for new physics, since post-CMB 
evolution and the formation of large scale structure is by no means 
a ``solved problem". Despite the strengths of linear gravitational 
collapse models in a Cold Dark Matter/Dark Energy 
($\Lambda _{\text{CDM}}$) universe~\cite{refWMAP}, there remain many 
loose ends, such as: the cuspy CDM halo problem, and the possible 
dearth of identifiable satellite-galaxy-sized structures in the local 
universe~\cite{refPrim}; the overabundance of high-\textit{z} clusters 
predicted by low-density $\Lambda _{\text{CDM}}$ models, implying 
actual values of $\Omega _{\text{M}}$ 
higher than those in the concordance model~\cite{refXMM1, refXMM2}; 
the unexpected existence of very massive galaxies at 
high-\textit{z}~\cite{refGenz}; the troubling, persistent tendency 
of Type Ia SNe data to indicate best-fit values of 
$\Omega _{\text{tot}} > 1$~\cite{refBarris} and/or (for Dark Energy) 
$w < -1$; the unknown behavior and composition of the Dark Energy 
itself (not to mention the Dark Matter)~\cite{refTurner1}; the 
possibility of non-Gaussian and hemisphere-asymmetric behavior in 
the WMAP data~\cite{refNonGauss}; and the question of what is 
causing the lack of low-order multipole power in the CMB, as well 
as its rolling spectral index of fluctuations~\cite{refWMAP}.

Searching for meaningful anisotropy in the cosmic expansion is not 
without empirical justification. Well documented are the large 
discrepancies that have been historically found between different 
measurements of the Hubble Constant, yielding a broad and non-Gaussian 
distribution of results~\cite{refCGR}; and though the error bars on 
$H_{0}$ have been greatly reduced over the years, even results 
quoted as demonstrating concordance (e.g., the agreement between 
values for $H_{0}$ found by WMAP and by the Hubble Key 
Project~\cite{refWMAP}), remain tempered by the fact that different 
types of $H_{0}$ measurements (e.g., combining S-Z effect and 
cluster X-ray flux measurements) still give somewhat discordant 
values~\cite{refWMAP}. In addition, in the high-\textit{z} supernova 
data used to prove cosmic acceleration, though the accelerating trend 
is statistically strong~\cite{refRiess1}, there remain enormous 
fluctuations for the individual SNe scattered about that trend (to be 
shown below). While the bulk of such discrepancies are doubtless due 
to a number of factors unrelated to the expansion itself -- e.g., 
systematic errors (theoretical and observational), poorly-understood 
physical processes causing variations in SN luminosity~\cite{refSNvar}, 
measurement difficulties leading to very large statistical errors, 
scarcity of data, etc. -- it remains possible that at least 
some of this SN scatter (and some of the disagreement between 
different $H_{0}$ measurements) represents real physics, and appears 
due to unknown dependencies of the expansion rate on angular position 
in the sky, which has been virtually ignored up to now by cosmological 
research because of the assumption of isotropy.

Despite the broad assumption (and validity) of the Cosmological 
Principle in general, we are not the first to explore the possibility 
of meaningful cosmic anisotropy on the largest observable scales. 
Zehavi \textit{et al.}~\cite{refHubbBubb} searched the early 
high-\textit{z} supernova data (with marginal, positive results) 
for possible evidence that we might live in a local ``bubble" of 
faster Hubble flow within a shell bounded by the local Great Walls. 
Alternatively, a detailed examination of the question by 
Lahav~\cite{refOFRW} led to some evidence of overall homogeneity, 
such as: agreement in shape and amplitude (except at the largest 
scales) between the 2dFGRS spectrum and a $\Lambda _{\text{CDM}}$ 
linear-regime perturbations model~\cite{ref2dFOL}; the near-convergence 
between the CMB dipole and the IRAS galaxy clustering 
dipole~\cite{refOFRW, ref2MASSdipole}; some evidence (though 
conflicting) of an isotropic distribution of very distant radio 
sources; an upper limit (produced using substantial theoretical 
interpretation) on anisotopy of sources contributing to the X-ray 
background; the apparent absence of big voids in the Lyman-a forest 
(covering $1.8 < z < 4$, roughly); and, anisotropy constraints by 
Kolatt and Lahav~\cite{refKOsne} from Type Ia SNe. (Somewhat 
curious is this last result, in which Kolatt and Lahav interpret a 
mild rejection of isotropy at the $\sim 70\%-80\%$ confidence level 
as evidence \textit{for} FRW behavior at these scales.) In short, 
there are a number of lines of evidence pointing towards isotropy 
and homogeneity on very large (sub-CMB) scales, though none of them 
are completely convincing at this time. The strongest reason for 
believing in extended FRW behavior long after the CMB epoch is still 
mainly a drive for concordance -- a theoretical motivation, not 
empirical proof.

Direct probes of cosmic structure made by mapping the universe still 
remain inconclusive in demonstrating large-scale homogeneity. Despite 
perennial expectations of reaching scales large enough for which 
structure finally gives way to smoothness, evidence continues to be 
found for apparently real structure at ever-increasing scales. 
Examples include: the large ``Local Hole", a significant deficit of 
galaxies in the APM survey area, re-verified with 2MASS data, and 
implying possible non-Gaussian clustering on scales up to 
$\sim 300 h^{-1}$ Mpc~\cite{refFrith}; $\sim 200-300$ Mpc sized 
structure detected in the 2dF QSO redshift survey~\cite{ref2dFQSO}; 
and the SDSS detection of the gigantic ``Sloan Great Wall", a 
structure 80\% larger than the CfA Great Wall (450 Mpc wide) in 
comoving coordinates~\cite{refUmap}.

The implications of such results are debated: Mueller and 
Maulbetsch~\cite{refSupSDSS} claim that in contrast to other, 
earlier results, the supercluster and void structure in the SDSS 
data is well reproduced by high-resolution $\Lambda _{\text{CDM}}$ 
simulations; Miller \textit{et al.}~\cite{ref2dFQSO} demonstrate 
(using a model requiring some bias of QSO distribution with respect 
to the Dark Matter distribution) that their 2dF QSO detection of very 
large structure does not provide any evidence of collapsed, non-linear 
structures on scales larger than 100 Mpc; and Gott 
\textit{et al.}~\cite{refUmap} claim (counterintuitively, it may seem) 
that the detection of the unprecedentedly large Sloan Great Wall 
provides support \textit{for} the expected approach to large-scale 
homogeneity (based upon an \textit{a posteriori} argument that the 
size of Sloan Great Wall, despite being much larger than any previously 
known ``single" structure, was not the largest possible structure that 
\textit{could have} fit in the SDSS survey). Nevertheless, the overall 
lesson still remains: the transition from CMB-era smoothness to the 
recent, clumpy universe is still very poorly understood, and the 
potential importance of anisotropy and inhomogeneity in the 
``Dark-Energy-dominated" epoch of cosmic evolution continues to be 
an open question.

The immediate impetus for this analysis is the publication of a 
combined, standardized list of 230 Type Ia Supernovae~\cite{refHzSN2}, 
containing extinction, redshift, luminosity distance and angular sky 
position data. Similarly to Kolatt and Lahav~\cite{refKOsne}, we 
search the SN data for statistical evidence of a lack of uniformity 
in the Hubble flow. What is different, however, is that we do 
not interpret our results according to any particular cosmological 
model. While they express their results in terms of variations in 
$H_{0}$, $\Omega _{\text{M}}$, etc., we instead ask a bare 
statistical question: ``Have some regions of the universe been 
expanding faster than others?", once the average Hubble evolution 
with respect to \textit{z} is (empirically) removed. This has two 
advantages: it keeps our analysis much more independent of 
theoretical assumptions; and we avoid the problem of having 
to dilute data from a limited, noisy sample by dividing the 
statistical power of the results among 4 four different model 
parameters.

One other aspect of our analysis is that we separately consider 
SNe at ``intermediate-\textit{z}" ($.01 < z < \sim .1-.2$), vs. 
``high-\textit{z}" ($z > \sim .1-.2)$. We use different types of 
analyses because the former is distributed more evenly, giving 
better sky coverage; though the latter is of more interest here, 
since the higher-\textit{z} SNe occupy more 
cosmologically-significant volumes of space, and correspond to 
look-back times closer to the onset of Hubble-flow acceleration.

Finally, we note that this conference proceedings paper is just 
a general overview of our results; we plan to present a more 
detailed discussion of our statistical methods and results in a 
future journal article.

\section{Preparation of Data for Analysis}

The data analyzed here is taken from the work of 
Tonry \textit{et al.}~\cite{refHzSN2}, a compilation of 
Type Ia SN data from many sources, including the 
High-\textit{z} Supernova Search Team (HZT), the Supernova 
Cosmology Project (SCP), and other researchers throughout 
the years. The total sample includes 230 SNe, presented 
in as uniform a fashion as possible, with a common 
calibration. 

For most of our cosmological analysis, we apply the same 
data cuts as in \cite{refHzSN2}: we remove all SNe with 
$z < 0.01$ (to minimize scatter from non-Hubble-flow peculiar 
motions), and we remove SNe known to be heavily extinguished  
(i.e., $\textit{A}_{V} < 0.5$ mag, for SNe with listed 
$\textit{A}_{V}$ values). These cuts reduce the sample 
set from 230 to 172, but successfully eliminate many dramatic 
(and likely spurious) outliers from the cosmic expansion rate 
fits. Later, we will divide the data into groups of 
``mid-\textit{z}" and ``high-\textit{z}" SNe; but for now 
we consider the entire sample.

The distance modulus (in Mpc) for each supernova is given 
as \cite{refWeinberg}: 
\begin{equation}
(m - M) = 25 + 5~\text{log}(d_{L}) = 25 + 5~\text{log}(d_{L} H_{0}) 
- 5~\text{log}(H_{0})~, 
\label{dispeq1}
\end{equation}
where $d_{L}$ is the supernova's luminosity distance, 
and lists of log$(d_{L} H_{0})$ for all SNe are given 
in \cite{refHzSN2}.

The theoretical luminosity distance in an empty or 
coasting (i.e., $q_{0}=0$) universe would be 
\cite{refWeinberg}:
\begin{equation}
D_{L} = \frac{cz}{H_{0}} (1+\frac{z}{2}) ~,
\label{dispeq2}
\end{equation}
with an expression similar to Eq.~\ref{dispeq1} for 
calculating the theoretical distance modulus, 
$(m - M)_{coasting}$. Any cosmic acceleration or 
deceleration is then found by computing 
the residuals, as follows (for the $i^{th}$ SN):
\begin{widetext}
\begin{equation}
\Delta(m - M)_{i} \equiv \Delta _{i} = [(m - M) - 
(m - M)_{coasting}]_{i} = 5~[\text{log}(d_{L} H_{0}) - 
\text{log}(c z) - \text{log}(1+\frac{z}{2})]_{i} ~.
\label{dispeq3}
\end{equation}
\end{widetext}

In most analyses, these $\Delta _ {i}$'s would then 
be used to find a best-fit cosmological model to be 
interpreted in terms of $\Omega _{m}$, 
$\Omega _{\Lambda}$, etc. But what we are interested 
in here is not finding best-fit model parameters, but 
in studying the \textit{scatter} of the data around 
the average cosmological expansion (whatever that is, 
assuming it to be well defined), for potential signs 
of cosmic anisotropy and inhomogeneity. To do this 
in a purely statistical, model-independent way, 
we perform a very simple fit to the data, and then 
subtract this fit, $f(z)$, from the $\Delta _ {i}$'s 
to compute the ``modified residuals": 
\begin{equation}
\delta (m - M)_{i} \equiv \delta _{i} = 
[\Delta(m-M)-f(z)]_{i} ~.
\label{dispeq4} 
\end{equation}
These $\delta _ {i}$'s are reasonably independent 
of \textit{z}, and can thus be used for SN statistical 
analysis.

\begin{figure}
\includegraphics[width=82mm]{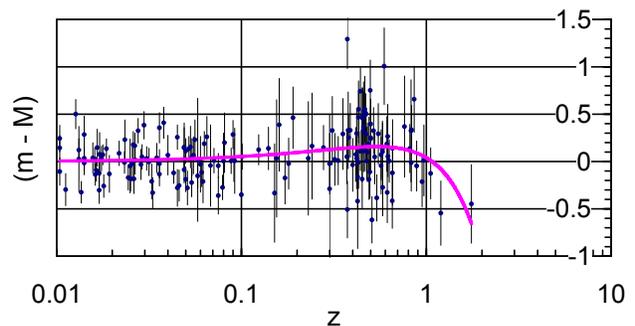}
\caption{\label{Fig1}Residuals of $\Delta(m-M)$ for Supernovae, 
plot vs. best-fit $2^{nd}$-order polynomial, 
$y = 0.586 z - 0.547 z^{2}$.}
\end{figure}

A polynomial fit to the $\delta _{i}$ for the 172 
SNe is shown in Figure \ref{Fig1}, with a best-fit 
function of $f(z) = (0.586 z - 0.547 z^{2})$. 
The essential physics is fairly well modeled (for this 
limited range of \textit{z}) with just a 
2$^{\text{nd}}$-order polynomial, showing both the 
current cosmic acceleration and the earlier deceleration. 
This fit yields a $\chi^{2}$ of 230; a little bit high, 
but good enough for our purposes. The cosmological 
fits in \cite{refHzSN2} do somewhat better, but they 
achieve that partially by artificially enlarging the 
error bars (which are somewhat heuristically generated 
in the first place) to include a velocity uncertainty 
(or dispersion) of 500 km/sec. We do not do this here, 
since such a procedure actually obscures the effect 
which we are trying to study (i.e., their ``noise" is 
our data).

\begin{figure}
\includegraphics[width=82mm]{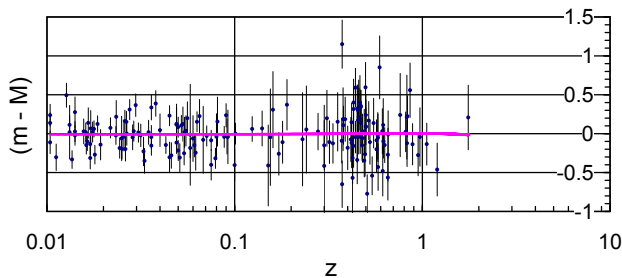}
\caption{\label{Fig2}Modified residuals, 
$\delta(m-M)$, from Fig.~\ref{Fig1} fit.}
\end{figure}

For any fit with a reasonable number of parameters, 
however, the SNe will always have tremendous scatter 
about the best-fit trendline. In this case, the index 
of fit~\cite{refRossStat} \textit{R} is only 
0.32, indicating that $\sim$70\% of the variation 
in the data, roughly speaking, has nothing to do with 
this (or any similar) fit. Tonry 
\textit{et al.}~\cite{refHzSN2} note the extreme 
noisiness of this data, and deal with it by binning 
the SNe, and taking medians over each redshift bin. 
While this may be useful for the analysis of 
cosmological models, it once again is a procedure 
which masks the physics that we wish to study here.

A plot of our fit-removed $\delta _{i}$'s is shown in 
Fig.~\ref{Fig2}, depicting a similar-looking (though
now visibly random) scatter. This is the main data 
that we analyze in this paper.

\section{Presentation of Results}

Here we give a brief overview of some results of our 
Supernova analysis. 

\begin{figure}
\includegraphics[width=82mm]{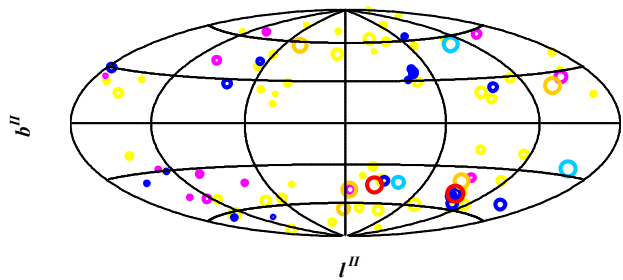}
\caption{\label{Fig3}Hammer-Aitoff plot of $\delta(m-M)$ for SNe at 
$.01<z<.2$ (with bubble size $\propto z$), color-coded depending upon 
$\delta$ (blue for low values, yellow for medium values, red/pink for 
high values).}
\end{figure}

\begin{figure}
\includegraphics[width=82mm]{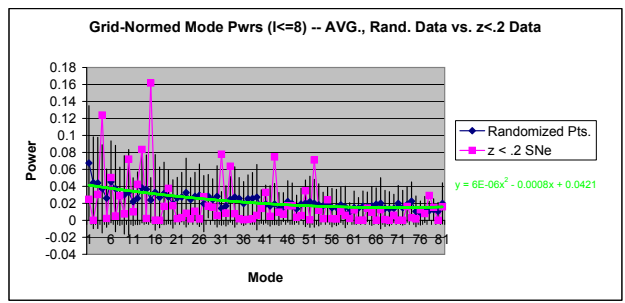}
\caption{\label{Fig4}Spherical harmonic modal decomposition of 
$\delta(m-M)$ for $.01<z<.2$ SNe (pink), vs. modal decompositions 
for 300 simulated random skies (blue).}
\end{figure}

First, considering the ``mid-\textit{z}" 
(i.e., $.01 < z < \sim .1-.2$) SNe data, 
we divide the SNe into 3 groups, depending 
upon their $\delta _{i}$ values: ``high", ``mid", 
or ``low". A plot of these data on the sky is shown 
in Fig.~\ref{Fig3}; and a modal analysis (using 
Spherical Harmonic modes) is depicted in Fig.~\ref{Fig4}, 
shown along with the modal decompositions of 300 randomly 
simulated skies, for comparison.

\begin{figure}
\includegraphics[width=82mm]{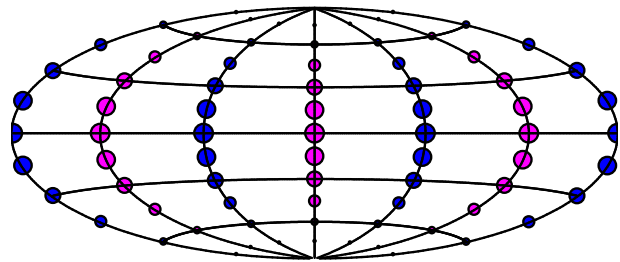}
\caption{\label{Fig5}Single largest mode from the decomposition 
in Fig.~\ref{Fig4}, $Y_{33}Cosine$ (real mode, cosine phase).}
\end{figure}

Some significant results here: the possible detection of a 
moderate dipole (larger than 80\% of simulated skies); no evidence 
of a significant quadrupole mode (smaller than $\sim 70\%$ of 
simulated skies); and the possible detection of one or two especially 
large anisotropy modes, particularly the $Y_{33}$ mode (cosine phase). 
The structure of this specific mode is demonstrated in Fig.~\ref{Fig5}; 
and future analysis with more data will be crucial in determining 
whether this is a real sign of anisotropy in the cosmic expansion, or 
merely a random event from the decomposition of this data set into a 
large number of modes.

\begin{figure}
\includegraphics[width=82mm]{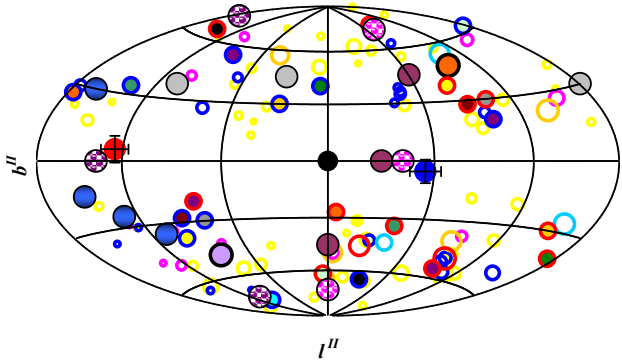}
\caption{\label{Fig6}Comparison of dipole direction from the 
decomposition in Figs.~\ref{Fig3} \& \ref{Fig4}, vs.~various 
galactic, intergalactic, and cosmological dipoles. 
\smallskip \smallskip}
\end{figure}

In an attempt to interpret the dipole mode present in the data 
(if it is real), Figure \ref{Fig6} compares the \textit{direction} 
of this dipole with a variety of other known, 
cosmologically-significant dipoles or significant directions -- 
such as the CMB Dipole, and dipole hotspots/coldspots from 2Mass, 
IRAS, Local Group velocity, etc.; and the Milky Way center, 
the Great Wall, the Great Attractor, the Perseus-Pisces Supercluster, 
the Supergalactic Plane, etc. Simple inspection does not indicate 
any clear, obvious alignment of the dipole we have derived from 
the SN data with any of these other dipoles, though it is possible 
that the Supergalactic Plane does run through the hot/coldspots of 
this apparent SN dipole. In any case, future supernova data should 
more precisely determine the direction and magnitude of the SN dipole 
(if it exists), and determine whether or not there is any real 
correlation with other cosmic dipoles.

\begin{figure}
\includegraphics[width=82mm]{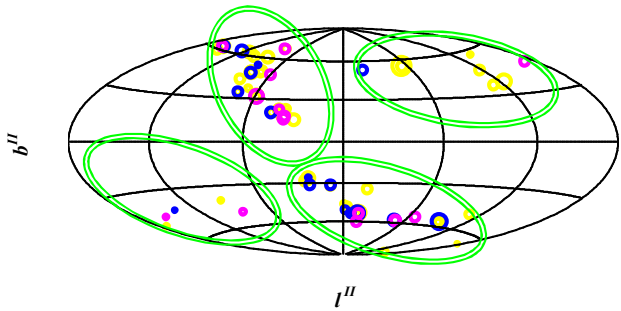}
\caption{\label{Fig7}Hammer-Aitoff plot of $\delta(m-M)$ for SNe 
at $z>.1$ (with bubble size $\propto z$), color-coded depending 
upon $\delta$ (blue for low values, yellow for medium values, 
red/pink for high values). Data partitioned into 4 regions 
reflecting groupings in space and $z$.}
\end{figure}

Now we consider the ``high-\textit{z}" 
(i.e., $z>.1$) SNe data, again dividing it up into 
three groups, with ``high", ``mid", and ``low" 
$\delta _{i}$ values. A plot of these data on the sky 
is shown in Fig.~\ref{Fig7}, with the sky roughly divided 
into 4 quadrants, for statistical comparison. (The kind of 
modal analysis done earlier for the intermediate-\textit{z} 
SN data is not as appropriate here, due to the 
lack of any real ``all-sky coverage" by the data.)

Statistical tests (t-tests~\cite{refRossStat}) between the 
upper-left and lower-right quadrants (the quadrants with the most 
and highest-\textit{z} SN data) show mild, positive evidence of 
an asymmetry between the quadrants, with an effective statistical 
significance of $\sim 1 - 1.8 \sigma$ (depending upon the 
specific data cuts used).

\begin{figure}
\includegraphics[width=82mm]{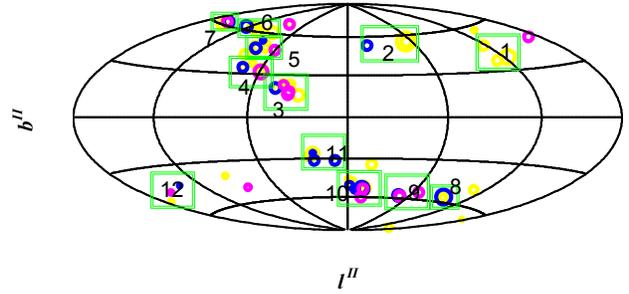}
\caption{\label{Fig8}Same data as in Fig.~\ref{Fig7}, 
re-partitioned into more smaller, more numerous groupings.}
\end{figure}

Lastly, as is shown in Fig.~\ref{Fig8}, we re-partition the 
SNe data into smaller, more specific groupings (partitioning 
done somewhat arbitrarily, but with the partitions being made 
as similar as is feasible in sky area). In this case, 
statistical tests (ANOVA tests~\cite{refRossStat}) 
again indicate mildly positive results ($\sim 1 - 2 \sigma$, 
depending upon the specific data cuts) for real differences 
between the $\delta _{i}$ values -- and thus the expansion 
rate histories -- for SNe in different parts of the sky.

\section{Conclusions}

We summarize these overall results as follows:

\smallskip

\smallskip

1. Given that Friedmann Robertson-Walker Cosmological Expansion 
depends upon the Cosmological Principle (Isotropy and Homogeneity), 
and that these assumptions break down in the recent universe (the 
Structure Forming and Accelerating Epoch), it is important to test 
the extent of this breakdown.

\smallskip

2. High-\textit{z} Supernovae are likely the best probes for these 
tests, as long as angular information is considered and analyzed, 
not just sky-position-averaged behavior as a function of \textit{z}.

\smallskip

3. For SNe at $.01 < z < .2$, we find some evidence of a dipole 
(which the Supergalactic Plane may be passing through), as well as a 
largest (real Spherical Harmonic) Anisotropy mode of $Y_{33}$Cosine. 
But both findings are difficult to quantify in terms of statistical 
significance.

\smallskip

4. For SNe at $z > \sim .2$, we find some mild positive evidence for 
a ``Dipole-like" Expansion Rate Anisotropy in opposite regions of the 
sky; and similar evidence for anistropies between smaller subdivided 
regions of the sky. But large gaps in sky coverage make these results 
hard to evaluate conclusively.

\smallskip

5. As more and better Supernova data are obtained (especially 
more all-sky coverage), we will be able to place more significant 
statistical limits on these potential anisotropies in the 
cosmological expansion.

\bigskip
\begin{acknowledgments}
We thank the High-\textit{z} Supernova Search Team and the 
Supernova Cosmology Project for making their data available 
in a standardized and easily accessible format which we have 
found useful for our analysis.
\end{acknowledgments}

\newpage

\bigskip

\end{document}